Coupled system description of perturbed KdV equation

Yair Zarmi

Jacob Blaustein Institutes for Desert Research
& Physics Department
Ben-Gurion University of the Negev
Midreshet Ben-Gurion, 84990
Israel

Abstract

In the multiple-soliton case, the freedom in the expansion of the solution of the perturbed KdV equation is exploited so as to transform the equation into a system of two equations: The (integrable) Normal Form for KdV-type solitons, which obey the usual infinity of KdV-conservation laws, and an auxiliary equation that describes the contribution of obstacles to asymptotic integrability, which arise from the second order onwards. The analysis has been carried through the third order in the expansion. Within that order, the solution of the auxiliary equation is a conserved quantity.

# 1. Introduction

The generic form of the KdV equation, perturbed through second order, is [1-6]:

$$\begin{aligned}
w_t &= 6ww_1 + w_3 \\
&+ \varepsilon\left(30\alpha_1 w^2 w_1 + 10\alpha_2 ww_3 + 20\alpha_3 w_1 w_2 + \alpha_4 w_5\right) \\
&+ \varepsilon^2 \begin{pmatrix} 140\beta_1 w^3 w_1 + 70\beta_2 w^2 w_3 + 280\beta_3 ww_1 w_2 + 14\beta_4 ww_5 \\ + 70\beta_5 w_x^3 + 42\beta_6 w_1 w_4 + 70\beta_7 w_2 w_3 + \beta_8 w_7 \end{pmatrix} + O(\varepsilon^2)
\end{aligned}$$
(1)

$$\left(|\varepsilon| \ll 1, \quad w_p \equiv \partial_x^p w\right)$$

One expands $w$ in powers of $\varepsilon$,

$$w(t,x) = u(t,x) + \varepsilon u^{(1)}(t,x) + \varepsilon^2 u^{(2)}(t,x) + O(\varepsilon^3) , \qquad (2)$$

Eq. (1) is integrable through $O(\varepsilon)$ [1-6]. Namely, if one terminates the analysis at $O(\varepsilon)$, then the zero-order approximation, $u$, is determined by a Normal Form that is integrable,

$$\begin{aligned} u_t &= 6uu_1 + u_3 + \varepsilon\, \alpha_4 S_3[u] + O(\varepsilon^2) \\ (S_3[u] &= 30u^2 u_1 + 10uu_3 + 20u_1 u_2 + u_5) \end{aligned}, \qquad (3)$$

and the first-order correction, $u^{(1)}$, has a closed-form expression as a differential polynomial in $u$:

$$u^{(1)} = a_1 u_2 + a_2 u_1 q + a_3 u^2 \qquad \left(q(t,x) = \partial_x^{-1} u(t,x)\right) , \qquad (4)$$

where

$$a_1 = -\tfrac{5}{2}\alpha_1 + \tfrac{5}{3}\alpha_3 + \tfrac{5}{6}\alpha_4 , \quad a_2 = -\tfrac{10}{3}\alpha_2 + \tfrac{10}{3}\alpha_4 , \quad a_3 = -5\alpha_1 + \tfrac{5}{3}\alpha_2 + \tfrac{10}{3}\alpha_4 . \qquad (5)$$

In Eq. (3), $S_3[u]$ is a symmetry of the KdV equation [1-9].

Eq. (3) has the same single- and multiple-soliton solutions as the unperturbed KdV equation. Denoting the wave number of a soliton by $k_i$, the only effect of Eq. (3) is to update the velocity of each soliton according to

$$v_i = 4k_i^2 + \varepsilon\alpha_4\left(4k_i^2\right)^2 + O(\varepsilon^2) \ . \tag{6}$$

However, this scheme cannot be extended to $O(\varepsilon^2)$, unless [2-6]

$$\begin{aligned}\mu = \tfrac{5}{3}\bigl(&3\alpha_1\alpha_2 + 4\alpha_2^2 - 18\alpha_1\alpha_3 + 60\alpha_2\alpha_3 - 24\alpha_3^2 + 18\alpha_1\alpha_4 - 67\alpha_2\alpha_4 + 24\alpha_4^2\bigr) \\ &+ 7\bigl(3\beta_1 - 4\beta_2 - 18\beta_3 + 17\beta_4 + 12\beta_5 - 18\beta_6 + 12\beta_7 - 4\beta_8\bigr) = 0\end{aligned} \tag{6}$$

If Eq. (6) is satisfied, then $u^{(2)}$, the second-order correction in Eq. (2), can be also solved for in closed form as a differential polynomial in the zero-order approximation, $u$, and the Normal Form Eq. (3) is updated through $O(\varepsilon^2)$ into:

$$u_t = 6uu_1 + u_3 + \varepsilon\,\alpha_4\,S_3[u] + \varepsilon^2\,\beta_8\,S_4[u] + O(\varepsilon^2) \ . \tag{7}$$

Eq. (7) has the same soliton solutions as the unperturbed KdV equation, with the velocity of each soliton now updated according to

$$v_i = 4k_i^2 + \varepsilon\alpha_4\left(4k_i^2\right)^2 + \varepsilon^2\beta_8\left(4k_i^2\right)^3 + O(\varepsilon^3) \ . \tag{8}$$

However, if $\mu \neq 0$, then the requirement that $u^{(2)}$ be a differential polynomial in $u$, spoils the integrability of the Norma Form [2-5]. Instead of Eq. (7), $u$ obeys the following equation:

$$\begin{aligned}&u_t = 6uu_1 + u_3 + \varepsilon\,\alpha_4\,S_3[u] + \varepsilon^2\left(\beta_8\,S_4[u] + \mu R^{(2)}[u]\right) + O(\varepsilon^2) \\ &\bigl(S_4[u] = 140u^3 u_1 + 70u^2 u_3 + 280uu_1 u_2 + 14uu_5 + 70u_1^3 + 42u_1 u_4 + 70u_2 u_3 + u_7\bigr)\end{aligned} \tag{9}$$

In Eq. (9), $S_4[u]$ is the next symmetry of the KdV equation [1-8] and $R^{(2)}[u]$ is the second-order obstacle to asymptotic integrability [2-5]. Whereas the value of $\mu$ is unique, the structure of $R^{(2)}[u]$ is not, owing to the freedom inherent in the expansion scheme.

The obstacle, $R^{(2)}[u]$, is not a symmetry of the KdV equation. Therefore, it spoils the integrability of Eq. (9). As a result, soliton parameters in the zero-order term, $u$, develop higher-order time dependence, non-KdV solitons are generated in $u$, and the elastic scattering picture of soliton collisions is lost in $u$ [3-5].

The difficulties reviewed above may be interpreted differently: That whereas $u^{(1)}$, the first-order term in Eq. (2), can be constructed as a differential polynomial in the zero-order term, $u^{(2)}(t,x)$, the second-order correction - may not [10, 11]. When $\mu \neq 0$, one must allow for a non-polynomial term in $u^{(2)}(t,x)$, and write it as

$$u^{(2)}(t,x) = \tilde{u}^{(2)}[u] + \eta^{(2)}(t,x) \quad . \tag{10}$$

In Eq. (9), $\tilde{u}^{(2)}[u]$ is the differential-polynomial part, and $\eta^{(2)}(t,x)$ is the non-polynomial part. The effect of the obstacle to integrability, $R^{(2)}[u]$, is accounted for by $\eta^{(2)}(t,x)$, and the integrable Normal Form, Eq. (8) is recovered.

The most general expression for $\tilde{u}^{(2)}[u]$ that is localized along soliton trajectories is

$$\begin{aligned}\tilde{u}^{(2)} &= b_1 u_4 + b_2 u_3 q + b_3 u_2 q^2 + b_4 u_1 q^3 + b_5 u_1 q^{(3)} + b_6 u q^4 + b_7 u q q^{(3)} \\ &+ b_8 u q^{(4)} + b_9 u u_2 + b_{10} u_1^2 + b_{11} u u_1 q + b_{12} u^2 q^2 + b_{13} u^3 \\ &\left(q = \partial_x^{-1}(u) \quad , \quad q^{(3)} = \partial_x^{-1}(u^2) \quad , \quad q^{(4)} = \partial_x^{-1}(u^2 q)\right)\end{aligned} \tag{11}$$

There is ample freedom in the choice of $b_k$, $1 \leq k \leq 13$. The Normal Form, Eq. (8), is recovered and the dynamical equation for $\eta^{(2)}(t,x)$ has exceptional characteristics with the following choice:

$$\begin{aligned}b_1 = \tfrac{5}{72}&\left(\begin{array}{l}135\alpha_1^2 - 24\alpha_1\alpha_2 - 12\alpha_2^2 - 36\alpha_1\alpha_3 - 200\alpha_2\alpha_3 + 92\alpha_3^2 \\ -114\alpha_1\alpha_4 + 216\alpha_2\alpha_4 + 20\alpha_3\alpha_4 - 77\alpha_4^2\end{array}\right) , \\ &-\tfrac{7}{6}(6\beta_1 - 3\beta_2 - 16\beta_3 + 14\beta_4 + 9\beta_5 - 15\beta_6 + 9\beta_7 - 4\beta_8)\end{aligned} \tag{12}$$

$$b_2 = \tfrac{25}{9}\left(3\alpha_1\alpha_2 - 2\alpha_2\alpha_3 - 3\alpha_1\alpha_4 + \alpha_2\alpha_4 + 2\alpha_3\alpha_4\right) - \tfrac{14}{3}(\beta_4 - \beta_8) \;, \tag{13}$$

$$b_3 = \tfrac{50}{9}(\alpha_2 - a_4)^2 \;, \tag{14}$$

$$\begin{aligned}b_5 = &\tfrac{5}{3}\left(9\alpha_1\alpha_2 + 2\alpha_2^2 + 6\alpha_1\alpha_3 - 20\alpha_2\alpha_3 + 8\alpha_3^2 - 6\alpha_1\alpha_4 - \alpha_2\alpha_4 + 2\alpha_4^2\right) \\ &- 7(\beta_1 + 2\beta_2 - 6\beta_3 + \beta_4 + 4\beta_5 - 6\beta_6 + 4\beta_7)\end{aligned} \;, \tag{15}$$

$$\begin{aligned}b_9 = &\tfrac{5}{9}\begin{pmatrix}135\alpha_1^2 - 39\alpha_1\alpha_2 - 12\alpha_2^2 - 6\alpha_1\alpha_3 - 190\alpha_2\alpha_3 + 72\alpha_3^2 - 129\alpha_1\alpha_4 \\ + 221\alpha_2\alpha_4 + 40\alpha_3\alpha_4 - 92\alpha_4^2\end{pmatrix} \\ &- \tfrac{14}{3}(12\beta_1 - 6\beta_2 - 32\beta_3 + 27\beta_4 + 18\beta_5 - 27\beta_6 + 18\beta_7 - 10\beta_8)\end{aligned} \;, \tag{16}$$

$$\begin{aligned}b_{10} = &\tfrac{5}{9}\begin{pmatrix}90\alpha_1^2 - 3\alpha_1\alpha_2 - 9\alpha_2^2 - 12\alpha_1\alpha_3 - 160\alpha_2\alpha_3 + 64\alpha_3^2 \\ -93\alpha_1\alpha_4 + 162\alpha_2\alpha_4 + 30\alpha_3\alpha_4 - 69\alpha_4^2\end{pmatrix} \\ &- \tfrac{7}{3}(18\beta_1 - 9\beta_2 - 48\beta_3 + 43\beta_4 + 27\beta_5 - 48\beta_6 + 32\beta_7 - 15\beta_8)\end{aligned} \;, \tag{17}$$

$$b_{11} = \tfrac{100}{3}\alpha_1(\alpha_2 - \alpha_4) - 28(\beta_4 - \beta_8) \;, \tag{18}$$

$$\begin{aligned}b_{13} = &\tfrac{25}{9}\begin{pmatrix}18\alpha_1^2 - 9\alpha_1\alpha_2 - 2\alpha_2^2 + 6\alpha_1\alpha_3 - 20\alpha_2\alpha_3 + 8\alpha_3^2 \\ -18\alpha_1\alpha_4 + 25\alpha_2\alpha_4 - 8\alpha_4^2\end{pmatrix} \\ &- \tfrac{7}{3}(15\beta_1 - 10\beta_2 - 30\beta_3 + 27\beta_4 + 20\beta_5 - 30\beta_6 + 20\beta_7 - 12\beta_8)\end{aligned} \;, \tag{19}$$

$$b_4 = b_6 = b_7 = b_8 = b_{12} = 0 \;. \tag{20}$$

With this choice for $b_k$, the equation for $\eta^{(2)}(t,x)$ becomes in this order

$$\partial_t \eta^{(2)} = \partial_x \left\{ 6(u\eta^{(2)}) + \partial_x^2 \eta^{(2)} + \mu u(u^3 + uu_2 - u_1^2) \right\} \;. \tag{21}$$

If u is a multiple-soliton solution of Eq. (8), then a solution of Eq. (19), which is bounded for fixed $t$, obeys the conservation law

$$\frac{d}{dt}\int_{-\infty}^{+\infty} \eta^{(2)}[u]\,dx = 0 \;. \tag{22}$$

If this is the case, then the perturbed KdV equation has been transformed by the perturbation scheme described above into a system of two equations: The Normal Form for ordinary solitons, which obey the well-know infinity of conservation laws [1-9] and Eq. (21) for the effect of the obstacle to asymptotic integrability. The latter generates a conserved quantity (at least in this order of the expansion).

That this is the indeed case is seen as follows. The differential polynomial $u(u^3 + uu_2 - u_1^2)$ in Eq. (21) is a *local special polynomial*. It vanishes identically if the single-soliton solution,

$$u_{Single}(t,x;k) = 2k^2 / \cosh[k(x+vt)] ,\qquad(23)$$

is substituted for $u$. As a result, $u(u^3 + uu_2 - u_1^2)$ is localized around the origin, and vanishes exponentially fast in all directions in the $x - t$ plane, if computed for a multiple-soliton solution [10, 11]. Hence, if $\eta^{(2)}(t,x)$ is bounded, then for fixed $t$, Eq. (22) is obeyed.

That $\eta^{(2)}(t,x)$ is bounded, and, in fact vanishes as $|x| \to \infty$ for fixed $t$, is seen as follows. Define

$$\eta^{(2)} = \mu \partial_x \omega^{(2)} .\qquad(24)$$

If $\eta^{(2)}(t,x)$ is bounded, then the equation for $\omega^{(2)}$ is

$$\partial_x \omega^{(2)} = 6u \partial_x \omega^{(2)} + \partial_x^3 \omega^{(2)} + u(u^3 + uu_2 - u_1^2) .\qquad(25)$$

The solution for $\omega^{(2)}[u]$ is bounded because the driving term in Eq. (25) does not resonate with the homogeneous part of the equation [11] and vanishes exponentially fast in all directions in the $x - t$ plane. As an example, Eq. (25) was solved numerically for zero-initial data at a large negative value of $t$ and vanishing boundary values for $x$, when $u$ is a two-soliton solution of the KdV equation, with soliton wave numbers equal to 0.1 and 0.2. Fig. 1 shows the solution for $\omega^{(2)}[u]$. It is comprised of a soliton and an anti-soliton, accompanied by a decaying dispersive wave.

Within the numerical accuracy, the soliton and anti-soliton have the same parameters (velocities, wave numbers and phase shifts) as the zero-order solitons in $u$. Up to overall amplitudes, determined from the numerical solution, the soliton and the anti-soliton are indistinguishable from the ordinary single-KdV solitons. The dispersive wave had been found in previous numerical works [12-14]. This work identifies the specific term that generates it.

With $\omega^{(2)}[u]$ bounded (in fact, vanishing as $|x| \to \infty$ for fixed $t$), $\eta^{(2)}(t,x)$, clearly, obeys the conservation law, Eq. (22), to lowest order.

To extend Eqs. (21) and (22) to $O(\varepsilon)$, requires that a third-order perturbation is appended to Eq. (1), and that the series for $w$ is computed through third order. This has been performed. The freedom in the expansion allows one to update Eq. (21) so that the result is also conservative (i.e., the right-hand side is a complete differential with respect to $x$):

$$\partial_t \eta^{(2)} = \partial_x \left\{ 6\left(u\eta^{(2)}\right) + \partial_x^2 \eta^{(2)} + \mu u\left(u^3 + uu_2 - u_1^2\right) + \varepsilon\left(A\left[u,\eta^{(2)}\right] + P[u]\right)\right\} \quad . \tag{26}$$

$A[u,\eta^{(2)}]$ is a differential polynomial in $u$ and $\eta^{(2)}$. It is linear in $\eta^{(2)}$. (Only the fourth-order analysis will generate in the extended version of Eq. (21) terms that will be quadratic in $\eta^{(2)}$.) Moreover, thanks to the freedom in the expansion, the driving term, $P[u]$, can be shaped so that it is also a local special differential polynomial in the zero-order term, $u$. Namely, it vanishes identically when computed for a single-soliton solution, and, as a result, is localized around the origin and falls of exponentially fast in all directions in the $x - t$ plane when $u$ is a multiple-soliton solution. Consequently, the conclusions of the lower-order analysis can be extended to the next order, and the validity of Eq. (22) is extended to, at least, $O(\varepsilon)$. The detailed results and the motivation that leads to the choice of coefficients (Eqs. (12-20)) will be published in a full-size paper.

FIGURE CAPTIONS

Fig. 1 Wave driven by second-order obstacle to asymptotic integrability (Eq. (25)).

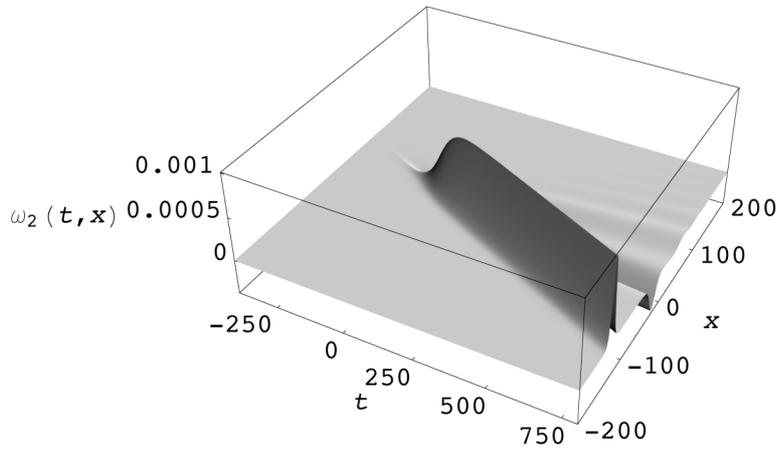